# Pocket Game Jams: a Constructionist Approach at Schools

**Anja Petri, Christian Schindler, Wolfgang Slany, Bernadette Spieler**
Graz University of Technology
{apetri, cschindler, wslany, bspieler}@ist.tugraz.at

**Jonathan Smith**
GameCity and the UK National Videogame Arcade
jonathan@gamecity.org

**ABSTRACT**

The constructionist approach is more interested in constructing personal experience than about acquiring information. It states that learning is most effective when building knowledge through active engagement. Experiential and discovery learning by challenges inspire creativity, and projects allow independent thinking and new ways of learning information. This paper describes how the "No One Left Behind" (NOLB) project plans to integrate this approach into school curricula using two concepts. The first one is to enable students to create their own games with Pocket Code by using its easy-to-learn visual programming language. The second concept is to foster collaboration and teamwork through hands-on sessions by conducting Game Jams using Pocket Code, so called Pocket Game Jams. We present insights into such a Pocket Game Jam and give an outlook on how we will use this concept.

**ACM Classification Keywords**

K.3.1 **[Computer Uses in Education]**: Collaborative learning; K.3.2 **[Computer and Information Science Education]**: Computer science education, Curriculum

## INTRODUCTION

Psychologists and pedagogues following the constructionist approach state three main goals. First to rethink traditional education without a step-by-step guidance and to imagine new environments [1], second to allow students to engage in meaningful and relevant problem-solving activities, and third to integrate new tools, media, and technologies in school lessons [8]. In this context, in the "No One Left Behind" (NOLB) project we strive to include digital game-based learning into school curricula in the interest of students aged from 12 to 17. We will combine two concepts into school lessons.

First, we will use digital game technologies [7] within the school curriculum. Smartphones and the use of apps are part of our culture and are changing the way in which many, particularly teenagers, act in social situations. An Australian survey of 1365 parents of smartphone owning children aged 3 to 17 shows that the kids spend an average of more than 21 hours per week using their devices [13]. This suggests that these children may see Pocket Code as a welcome addition to traditional teacher-centred classrooms and that parents may see pedagogical value in time their kids spend on their phones engaged in creative fun activities.

Second, we plan introducing Pocket Code to students by conducting Game Jams that are creative, exciting, and interesting social experiences [6]. The goal of a Game Jam is to design a game, together in teams, usually within 48 hours. The key elements are time pressure and a given theme. The deadline forces participants to be fast, cut corners, think outside the box, and finish a game within a fixed time frame. The theme is often kept secret until the beginning of the Jam to ensure equal conditions and to encourage the creativity and problem solving skills by thinking about a fitting game idea. This paper describes how a lesson with Pocket Code could be structured so it follows the constructionist approach by performing Game Jams.

In this paper we begin by introducing the constructionist approach, introduce Pocket Code, and describe the goals of the "No One Left Behind" project. In Section 2 we explain the general concept of a Game Jam. We then describe our experiences using Pocket Code at the Global Game Jam at the UK's National Video Game Arcade in Section 3. Section 4 is concerned with our first Pocket Game Jam with students of the computer science teacher trainee program and defines important metrics. Finally we describe a framework for Pocket Game Jams at schools.

## CONSTRUCTIONISM, POCKET CODE, AND THE "NO ONE LEFT BEHIND" PROJECT

Within the NOLB project we examine how to attract, motivate, and engage students with content from an academic curriculum and at the same time supporting the learning process and providing an effective learning experience. This section shortly explains the term constructionism, the Pocket Code app, and the objectives of the NOLB project.

**Constructionism:** Seymour Papert's constructionism is based on Jean Piaget's constructivism and focuses on the construction of knowledge through experiential learning. Constructivism provides a framework for optimizing the learning progress at different levels of children's development [11]. Younger children create their own subjective reality, depending on their own experiences and suited to their current needs and possibilities. With 11 years children enhance their capability of abstract thinking and start to think about probabilities, associations, and analogies. Constructivism states that a.) teaching is always indirect (teachers take the role of a coach), b.) knowledge is experience and build by the learner, and c.) learners are considered to be central in the learning process. In contrast to Piaget, Papert focuses on how ideas can be formed and transformed when expressed through different media, actualized in particular contexts, and worked out by individual minds. The iterative process of self-directed learning underlines the concept that humans learn most effectively when they are actively involved in the learning process and build their own structures of knowledge. In this model, communication about their work, the process of learning itself with peers, teachers, and collaborators, is an indispensable part of a student's learning [9].

Papert described the huge potential bringing new technology into the classroom [10]. He figured out that students learn more efficiently if they could see a concrete result of their computing efforts. Therefore he co-invented the LOGO programming language in the late 1960s at MIT. Logo was designed to have a "low threshold and no ceiling": It is accessible to novices, including young children, and also supports complex explorations and sophisticated projects by experienced users [12]. He states that software enhanced learning provide contexts for dialogue and interaction within the classroom, the schools, and the community leading to the social construction of knowledge.

Due to that we determine the following key factors for a constructionist classroom: work in groups to approach problems and challenges in real world situations, encourage creative experimentation, and provide hands-on opportunities.

**Pocket Code:** Pocket Code is an application for mobile devices. This app allows teenagers to create their own games, animations, interactive music videos, and many types of other apps, directly on their smartphones or tablets. It uses a visual programming language and is developed by the free and open source project Catrobat [2]. Pocket Code's aim is to enable teenagers to creatively develop and share their own software online. The app is freely available on Google Play. Pocket Code is inspired by, but distinct from the Scratch programming language developed by the Lifelong Kindergarten Group at the MIT Media Lab. In contrast to Scratch, no traditional PC is required for using Pocket Code, and Pocket Code is able to access the mobile device's sensors (e.g., acceleration, compass, inclination, multitouch). Similar to Scratch, programs in Pocket Code are created by snapping together command bricks. The bricks are arranged in "scripts" which can run in parallel, thereby allowing concurrent execution. Broadcast messages are used to communicate between objects and to trigger execution of scripts that listen to certain broadcast messages. By means of this mechanism, sequential or parallel execution of scripts is possible, either within the same object or over object boundaries. Figure 1 shows the command bricks of a working compass that was programmed using Pocket Code. The compass needle is continuously pointing north.

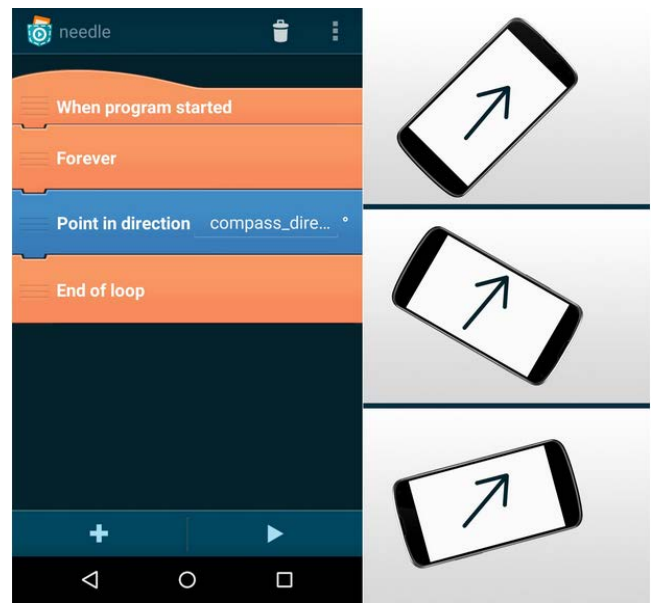

**Figure 1: Script of a compass created with Pocket Code.**

**The "No One Left Behind" Project:** The NOLB project validates its output conducting three pilot studies in Europe (Austria, UK and Spain) targeting 600 students between 12 to 17 years. Each pilot site will address a different social inclusion challenge: Gender exclusion, disability, and immigration. During the project we will create a new version of Pocket Code that integrates a set of game mechanics, dynamics, assets, and in-game analytics from leisure oriented digital games. Furthermore it will incorporate the academic curricula of different subjects at the piloting schools. In the future we will develop it to become an empowering tool that supports the constructionist approach and therefore the development of creativity, problem solving, logical thinking, system design, and collaboration skills.

**CONCEPT OF GAME JAMS AT UK'S NATIONAL VIDEOGAME ARCADE**

The National Videogame Arcade (NVA) was recently opened in the UK as a space to engage the public with videogames and game-making creativity, and operates regular Game Jams with both professional developers and school children [3]. The NVA uses Game Jams as an important tool in its mission to make game-making a universal skill, and to broaden and diversify the range of people able to express themselves as well as create new work through games. Supported by Nottingham Trent University, the NVA is a project run by GameCity, whose annual game festivals in Nottingham have established an international reputation, creating new opportunities for renowned developers to inspire and engage with a new generation of creators. An introduction from an established game developer, either in person or on video, is an important way to build the creative energy, which can drive Game Jam participants to exceed their expected limitations.

There are three main ways in which the NVA operates Game Jams:

1. With school groups as part of a pre-booked workshop session. These have been delivered to children between the ages of 7 and 16 "a very broad ability range" but in every case, the students have shown themselves able to conceptualize and execute new ideas to make their own games with original art, sounds, and game mechanics.

2. As part of "Game Club" sessions, which take place re- currently with children, aged 9 to 13.

3. With college-level developers, hobbyists, and independent creators, over the course of a weekend.

In each case, it is important that the results of the Jam are preserved and exhibited in order to foster a sense of achievement and pride, and to give a concrete outcome of the creative process which can then provoke feedback and self-reflection in order to build learning and progression. Because the NVA operates as a public space, GameCity can showcase these works by representing them in displays and exhibitions. They can also be shared and preserved digitally on a website.

The NVA was one venue for the Global Game Jam (GGJ) event that annually takes place in more than a hundred different countries all over the world, at the same time and with a common theme [4]. The Pocket Code team took part in this GGJ and created a game matching the GGJ's theme "What do we do now?" that also considered several diversifiers set forth by the organizers. We gained first insights into how to organize and divide work among a team. During the Jam we successfully developed a game with two very different levels and an engaging background story within 48 hours[1].

The idea of a Game Jam is to plan, design, and create games while working in teams. Pocket Code adheres to the constructionist approach and therefore provides a simple way of learning to program. So do Game Jams: They encourage working in a team, situational problem solving, and being creative.

**CONCEPT OF POCKET GAME JAMS**

Using the constructionist approach and combining it with the Game Jam idea, we try to enhance the factors collaboration and teamwork while using Pocket Code in Game Jams. We call this concept Pocket Game Jams (PGJ). During the GGJ we identified four challenges for future PGJs.

*Issue 1* - Working together in a team and merging of programs: To ensure teamwork and to program collaboratively, Pocket Code needs to be enhanced. Based on our experience by participating at the GGJ, we found out that it is difficult to work together on one program with Pocket Code, since the app does not support several persons working together in parallel. To support such cooperation, we plan to implement a backpack function to save objects, codes, looks, and sounds temporarily. As a result of our experience at the GGJ, as a first feature we added a "Merge into current program" function to combine two programs into one.

*Issue 2* – Effective division of work and appropriate team size: Our team at the GGJ consisted of three people: two programmers and one designer. While one was developing the first level of the game, the other one was creating the second level. The designer was taking care of the artwork, i.e., character design and background images. Based on this experience, we found that a successful PGJ team can be composed of three members. Thus a team could split up the tasks like we did and work efficiently on one game.

*Issue 3* - Suitable time frame: With 48 hours we had enough time for planning and programming our game. For conducting PGJs in a more formal school context there will most likely not be such a continuous and intensive period of 48 hours. Therefore we want to test what the minimal time frame is to create a game with Pocket Code. Obviously this will also depend upon previous experience with Pocket Code. We are planning to conduct PGJs within two, four, six, and 24 hours to answer the following questions: What time frame is suitable to conduct a PGJ for novice versus experienced participants? What is the minimum amount of time for a Pocket Game Jam? What is generally possible within each time frame? Note that in a school context, it might be easier to conduct Game Jams over a longer but less intensive period of time, where students would be able to work on games as a homework exercise over one or two weeks. We will also conduct this type of PGJs. As an example for a project we will provide a main theme e.g., gravity and a sub-theme for every group e.g., planets in case of a physics lesson.

---

[1] https://www.youtube.com/watch?v=F8y ZL5bcO0

*Issue 4* - Target group: As mentioned participants can be more or less experienced. Therefore we plan to conduct PGJs repeatedly with the same teachers and students over a longer period of time to study changes based on experience.

**POCKET GAME JAM: A FIRST ATTEMPT**

As we are planning to integrate Pocket Code in school curricula, we conducted our first Pocket Game Jam with university students of a computer science teacher training program to evaluate our first concept draft. This first PGJ should help us to adjust the previously described metrics, e.g., team size and time frame. The Jam was conducted on one afternoon in March 2015, and the time frame for the core part of the Jam was two hours. The PGJ was planned as follows:

4. Introduction to Pocket Game Jam (general information and installation)
5. Team-building, idea creation, and brainstorming
6. 10 minutes break
7. 2 hours for the core Pocket Game Jam
8. Merging and submission of the games
9. Presentation, feedback, and lessons learned

After the introduction we presented the theme for this PGJ - to create a learning game that has something to do with the students' other subject (each teacher in Austria has two subjects, e.g., Computer Science and History). Additionally the game jammers received a sheet with diversifiers to make the game design more exciting; such diversifiers represent a further challenge and could be used at the discretion of the participants. The slogan for diversifiers' states: Creativity is born from constraints [5].

The following diversifiers were suggested:

- The use of sensors: Acceleration, inclination, loudness, compass direction
- Integration of interactivity (e.g., when something is tapped...)
- Implementation of several game levels
- Testing the learning experience, e.g., through a quiz

Thus the students had to cope with two challenges. The first challenge was to think about possible ideas and include some diversifiers. The second challenge was to create a functional game within the limited amount of time. It was our task to support the teams during the brainstorming process. Helpful hints included that they should think first of the simplest thing that might possibly work (i.e., following an agile approach) that makes the game playable, and second to use only rough sketches instead of elaborated and well designed graphics. In order to achieve this, we encouraged them to draw a storyboard; a storyboard helps to order ideas and defines how the game will be played. After creating their storyboards they started programming. They also had to think of how to best split the work among team members. After nearly two hours all teams managed to successfully complete the creation of their first game level. During the whole PGJ the most interesting question for us was: What is possible within two hours?

The results showed that it is possible to create a game as a team and within a limited time frame of two hours. The games were executable, presented the general game idea, contained several diversifiers, and concepts for additional levels. All teams were very satisfied with the outcomes of their first Jam, but they mentioned that the time frame of two hours was too short. However, all participants agreed that they could imagine conducting PGJs within a school context (e.g., during a lesson) in the future, but they tagged it as "risky" to use during regular lessons. One solution could be to conduct such jams during a project day. Although programming with Pocket Code as a team was a new experience for everyone, they mentioned that it was very interesting and that it had worked smoothly when the tasks were split properly. The Jam is to be regarded as a preliminary study to evaluate the general concept of PGJs. We will need to conduct further Jams with our target group of students in order to collect meaningful data and to prove that this approach can be successfully adopted into school lessons.

**OUTLOOK: POCKET GAME JAMS AT SCHOOLS**

The idea of integrating Game Jams into school curricula is to make the initial contact with Pocket Code easier. It is not necessary to know exactly what variables, loops, or broadcast messages are, but it is important to experience why these principles are needed when programming a game. In the case of the NOLB project we are integrating Pocket Code not only in Computer Science lessons but also into courses like Spanish, Fine Arts, Music, or Physics. To integrate Pocket Code into non-computer science classes, it is necessary that teachers are open for using new technologies. To convince and to show them how easy it is to create games with Pocket Code, we want to conduct PGJs with teachers as participants. They will not have much more prior knowledge in using Pocket Code than their students, but we will follow the constructionist approach and let them create their own educational games in order to let them imagine how they could use PGJs in their classes as well.

In this paper we have described the benefits of a PGJ. In future we will encourage and support teachers to conduct PGJs with their students in order to introduce Pocket Code in their lessons and use it as a playful and engaging tool for both the Computer Science classes as well as other academic subjects. We are currently creating support material to help teachers to conduct PGJs on their own, especially for users with little or no knowledge of programming. Our ultimate goal is to make suitable material available online that is effectively motivating and supporting teachers so that they are empowered to organize their own Game Jams, first with their colleagues and then with their classes, without our direct involvement, in order to spread the PGJ idea to schools all over the world. Our previous described first attempt of a PGJ could be seen as an exploratory exercise to support further studies. We are at the beginning of the project and assume that PGJ represent a collaborative way to learning while having fun creating games in teams with Pocket Code.

Future work will include conducting introductory PGJs at the end of the school year (i.e. July 2015) within the schools' traditional project week, so that teachers have time over the holidays to think about possible adoption scenarios into their regular classes. We will also provide initial kickoff sessions for students before starting the Jams. The holidays are also ideal to contact us in case they need help developing their ideas and implement them in Pocket Code.

**ACKNOWLEDGMENTS**

This work has been partially funded by the EC H2020 Innovation Action No One Left Behind, Grant Agreement No. 645215.

BibTex entry:

```
@InProceedings{SPIELER2016ICGBL,
author = {Spieler, B . and Petri, A. and Slany, W. and Schindler, C. and Beltràn, M.E. and Boulton, H.},
title = {Pocket Game Jams: A Constructionist Approach at Schools},
booktitle = {n Proceedings of the 6th Irish Conference on game-Based Learning.},
series = {MobileHCI '15},
isbn = {978-1-4503-3653-6},
doi = {10.1145/2786567.2801610},
url = {http://doi.acm.org/10.1145/2786567.2801610},
publisher = {ACM},
location = {Copenhagen, Denmark},
month = {24-25 August, 2015},
year = {2015},
pages = {1207-1211}}
```